\documentclass[amsmath,amssymb]{revtex4}
\usepackage[dvips]{graphicx}
\usepackage{dcolumn}
\usepackage{bm}

\newcommand{\beq}{\begin{equation}}
\newcommand{\eeq}{\end{equation}}
\newcommand{\bea}{\begin{eqnarray}}
\newcommand{\eea}{\end{eqnarray}}

\begin{document}
\title[]
{Klein-Gordon equation for a particle in brane model}

\author{Sergey N. Andrianov}
\email{adrianovsn@mail.ru} \affiliation{Scientific center for
gravity wave studies ``Dulkyn'', Kazan, Russia}

\author{Rinat A. Daishev }
\email{Rinat.Daishev@ksu.ru} \affiliation{$^1$ Department of
Mathematics and Department of Physics, Kazan Federal University,
Kremlevskaya str. 18, Kazan 420008, Russia}

\author{Sergey M. Kozyrev}
\email{Sergey@tnpko.ru} \affiliation{Scientific center for gravity
wave studies ``Dulkyn'', Kazan, Russia}

\begin{abstract}\noindent
Brane model of universe is considered for a free particle. Conservation laws on the brane are obtained using the symmetry properties of the brane. Equation of  motion is derived for a particle using variation principle from these conservation laws. This equation has a form of Klein-Gordon equation. Comparison of squared Dirac-Fok-Ivanenko equation for a spin particle with Klein-Gordon equation in curved space has given an expression for chiral spin current variation through the derivative of spin connectivity. This chiral spin current is anomalous spin current corresponding to spontaneous chiral symmetry breaking of massive particle in the space of KG equation solutions. \\
\phantom{a}\\
PAC numbers: 04.20, 03.50
\end{abstract}

\maketitle \noindent

\section{Introduction}\noindent

There is Dirac-Fok-Ivanenko (DFI) equation \cite{Ivanenko}
describing fermions behavior being valid when the supersymmetry is
broken. Klein-Gordon (KG) equation for a particle describes
behavior of both bosons and fermions and is valid for the case of
global symmetry where currents analogous to Dirac chiral currents
are conserved  \cite{Mostafazadeh}. Squaring of DFI equation
yields KG equation in curved space \cite{Schrodinger, Shavokhina,
Weldon}. KG equation is also obtained from variation principle in
works \cite{Andrianov, Andrianov2, Pavsic}.

In present paper, we will derive starting from the symmetry
properties of the brane \cite{Papantonopoulos, Langlois, Brax} the
equation of motion for a particle in the framework of brane model.
This equation has a form of Klein-Gordon equation. We perform
comparison of squared DFI equation with KG equation containing
free term in concrete form corresponding to brane symmetry
approach. We derive by that an additional relation - the
expression for gamma five matrix through the derivative of spin
connection. Then we use it for determination of chiral spin
current manifesting chiral anomaly at spontaneous symmetry
breaking.


\section{ENERGY CONSERVATION LAW}
Let's consider our space as four dimensional hyper-surface that is
embedded in the space of higher dimension. Then interval for a
moving particle in normal Gauss coordinates can be written as

\begin{equation}
ds = \sqrt{g_{i j} d x^i d x^j-c^2 dt^2}, \label{eq1}
\end{equation}
where $g_{i j}$  is metric tensor, $ d x^i, d x^j $  are
differentials of coordinates ($i, j = 0,1,2,3$) on brane, $dt$ is
differential of universal time that is proportional to extra
dimensional coordinate ($ dx^4 = c dt$) and is the same for all
points on brane. Greek symbols will denote all indexes ($\alpha =
0,1,2,3,4$) including extra dimension.

We take the convention that the metric signature is ($+ - - - -$).
Then the action can be written as

\begin{eqnarray}\label{eq2}
S=mc   \int ^T_0 ds = \int ^T_0 L dt
\end{eqnarray}
where $m$ is mass of particle, $c$ is speed of light,

\begin{eqnarray}\label{eq3}
L = \sqrt{g_{i j} (\dot{m} x^i) (\dot{m} x^j) - m^2 c^2 },
\end{eqnarray}
is Lagrangian, $T$ is current value of universal time in
multidimensional space (proportional to the brane radius).

Let's introduce the symmetry of configuration space as single
parametric transformation group $f(q, \varepsilon)$:
\begin{subequations} \label{BDeqs}
\bea t &\rightarrow& t+\varepsilon,  x_i  \rightarrow x_i(t +
\varepsilon)   ,
\\
f(q_i, \epsilon)  &=& x_i(t + \varepsilon),  f(q_i, 0)  =
x_i(t)\eea
\end{subequations}

conserving Lagrangian  (\ref{eq3}). According to Noether's
theorem, first integral

\begin{equation}
I= \frac{\partial L}{\partial \dot{x}_i}h^i,  \label{eq4}
\end{equation}

where

\begin{equation}
h^i= \frac{\partial f^i}{\partial \varepsilon},  \label{eq4o}
\end{equation}
can be put in correspondence to each symmetry. Then
\begin{equation}
I= \frac{1}{L}g_{i j} m^2 \dot{x}^j \left( \frac{\partial x_i
(t+\varepsilon) }{\partial (t+\varepsilon)} \frac{\partial
(t+\varepsilon)}{\partial \varepsilon}\right)_{|\varepsilon=0},
\label{eq5}
\end{equation}
or
\begin{eqnarray}
g_{i j} m \dot{x}^i  m \dot{x}^j = const, \label{eq6}
\end{eqnarray}

If the particle moves uniformly and rectilinearly on the
background of Lorenz's metrics than we can choose reference system
where $\dot{x}^{(1)} = \dot{x}^{(2)}  = \dot{x}^{(3)} = 0$  and
$\dot{x}^{(0)} = 0$  when brane is expanding with velocity $c$.
Then (\ref{eq6}) yields
\begin{eqnarray}
g_{i j} m \dot{x}^i  m \dot{x}^j = m^2 c^2, \label{eq6o}
\end{eqnarray}

The same equation can be derived in the framework of quantum
mechanical treatment. The wave function of particle in
quasi-classical approximation is
\begin{eqnarray}
\psi &=& a e^{\frac{i S}{\hbar}}, \label{eq7}
\end{eqnarray}
where a is slowly varying amplitude, $S$ is action expressed by
formulas (\ref{eq2}, \ref{eq3}).   Let's differentiate both sides
of expression (\ref{eq7}) by $T$ neglecting the dependence of
amplitude on universal time
\begin{equation}\label{eq8}
\frac{\partial \psi}{\partial T}=a \frac{i}{\hbar}e^{\frac{i
S}{\hbar}}\frac{\partial S}{\partial T} = i
\frac{c}{\hbar}\sqrt{g_{i j} p^i p^j - m^2 c^2 \psi }
\end{equation}

where  $p^i = m x^i$.

If evolution of particle in brane does not depend on brane radius
then $\frac{\partial \psi}{\partial T}=\frac{\partial S}{\partial
T} = 0$  and
\begin{equation}\label{eq9}
g_{i j} p^i p^j = m^2 c^2.
\end{equation}

\section{KLEIN-GORDON EQUATION}

Expression (\ref{eq9}) can be rewritten in the following form for
wave function   in Hilbert space:

\begin{equation}
p_i p^i \psi =m^2 c^2 \psi. \label{eq10k}
\end{equation}

Let's consider functional variation of relation (\ref{eq10k})  in
the vicinity of $x$. Complete variation of momentum vector can be
written as the sum of functional variation $\delta p$  of vector
$p$ at the comparison of $p(x')$  with $p(x'')$   in the vicinity
of $p(x)$  at the parallel transfer of momentum vector in
universal space and ordinary variation  $dp$.  Then, it can be
written that
\begin{equation}
\triangle p=\frac{1}{2}(p\left( x^{\prime }\right) -p\left(
x^{\prime\prime} \right) )=\frac 12 (p\left( x^{\prime }\right)
-\stackrel{\sim }{p}\left( x^{\prime }\right) +\stackrel{\sim
}{p}\left( x^{\prime }\right) -p\left( x^{\prime\prime}\right))
=\frac 12\delta p+\frac 12 dp, \label{eq15k}
\end{equation}
where
\begin{equation}
\delta p=p^{\prime }\left( x^{\prime }\right) -\ \stackrel{\sim
}{p}\left( x^{\prime }\right)  \label{eq16k}
\end{equation}
and
\begin{equation}
dp=\ \stackrel{\sim }{p}\left( x^{\prime }\right) -p\left(
x^{\prime\prime}\right) , \label{eq17k}
\end{equation}
$\stackrel{\sim }{p}\left( x^{\prime }\right) \ $is momentum
vector at its parallel transfer in the universal space from point
$x^{\prime\prime}= x -\delta x  $ to point  $x^{\prime}= x +\delta
x  $. If trajectory of particle is geodetic one then

\begin{equation}
dp_i=\frac{\partial p_i}{\partial x^k}dx^k=0,  \label{eq18k}
\end{equation}

\begin{equation}
\delta p_i=\stackrel{\sim }{p}_k\Gamma _{i\alpha}^k\delta
x^\alpha, \label{eq19k}
\end{equation}
where $\delta x^\alpha=\frac{1}{2}(x^{\prime \alpha}-x''^\alpha)$.
Rome indexes numerate here coordinates of usual four-coordinate
space and Greek indexes numerate coordinates of universal
five-coordinate space. It was assumed at formulation of
 (\ref{eq19k}) that $\widetilde{p}_4=0 $.

 Then, it can be written, omitting stroked index
of momentum vector,

\begin{equation}
p\left( x^{\prime }\right) =p\left( x\right) + \frac12\delta p.
\label{eq20k}
\end{equation}

At the transform {\it x }$\rightarrow ${\it \ x'}, relation
(\ref{eq10k}) is transforming accounting (\ref{eq20k}) to the
following form:

\begin{equation}
\left\{ p_ip^i+\frac12 \left(p_i\delta p^i+\delta
p_ip^i\right)+\frac14\delta p_i\delta p^i \right\} \psi=m^2c^2
\psi. \label{eq21k}
\end{equation}

Let's pass in relation (\ref{eq21k}) to operators acting in
Hilbert space of wave functions $\psi \left( x\right) $. We
represent for this sake the components of vector {\it p} as

\begin{equation}
p_i=-i\hbar \frac \partial {\partial x^i},  \label{eq22k}
\end{equation}

and rewrite relation (\ref{eq19k}) as

\begin{equation}
\delta p_i=-i\hbar\left\{ \Gamma _{i\alpha }^k\delta x^\alpha
\right\} _{;k}, \label{eq23k}
\end{equation}
assuming that $ \widetilde{p_k}$  is a covariant derivative.

Let's consider the first term in the left side of equation
(\ref{eq21k}). For this purpose, we represent it in the form

\begin{equation}
p_ip^i \psi =p_ig^{ij}p_j \psi.  \label{eq24k}
\end{equation}

Using expression (\ref{eq22k}), we get

\begin{equation}
p_ip^i \psi=-\hbar ^2\left( \frac{\partial g^{ij}}{\partial
x^i}\frac
\partial {\partial x^j}+g^{ij}\frac{\partial ^2}{\partial
x^i\partial x^j}\right)\psi . \label{eq25k}
\end{equation}

Let's use well known relation

\begin{equation}
\frac{\partial g^{ij}}{\partial x^k}=-\Gamma _{mk}^ig^{mj}-\Gamma
_{mk}^jg^{im}.  \label{eq26k}
\end{equation}

Then

\begin{equation}
p_ip^i \psi=-\hbar ^2\left( g^{ij}\frac{\partial ^2}{\partial x^i\partial x^j}%
-g^{mj}\Gamma _{mi}^i\frac \partial {\partial x^j}-g^{im}\Gamma
_{mi}^j\frac
\partial {\partial x^i}\right)\psi .  \label{eq27k}
\end{equation}
Changing indexes of summation, we get

\begin{equation}
p_ip^i \psi=-\hbar ^2g^{ij}\left( \frac{\partial ^2}{\partial x^i\partial x^j}%
-\Gamma _{ij}^k\frac
\partial {\partial x^k}\right)\psi.  \label{eq28k}
\end{equation}
Let's consider second term in the left side of equation
(\ref{eq15k}), rewriting it in the form

\begin{equation}
p_i\delta p^i \psi=p_ig^{ij}\delta p_j \psi.  \label{eq29}
\end{equation}

Using formula (\ref{eq26k}), we get

\begin{equation}
p_i\delta p^i \psi =\left\{g^{ij}(p_i\delta p_j)+i\hbar \left(
g^{ij}\Gamma _{im}^m+g^{im}\Gamma _{im}^j\right) \delta p_j +
g^{ij}\delta p_j p_i \right \} \psi. \label{eq30}
\end{equation}

Let's write in its direct form the covariant derivative in the expression (%
\ref{eq23k}):

\begin{equation}
\delta p_j=-i\hbar \left(\frac{1}{2} \Gamma _{jk}^k +\frac{\partial \Gamma _{j\alpha}^k}{%
\partial x^k}\delta x^\alpha-\Gamma _{lk}^k\Gamma _{l\alpha}^k\delta x^\alpha+\Gamma
_{lk}^k\Gamma _{j\alpha}^l\delta x^\alpha\right) .  \label{eq31k}
\end{equation}

We get from formula (\ref{eq31k})
\begin{equation}
\delta p_j=-i\hbar \left( \frac12\Gamma _{jk}^k+R_{j \alpha}\delta x^\alpha+\frac{\partial \Gamma _{jk}^k}{%
\partial x^\alpha}\delta x^\alpha\right) .  \label{eq32k}
\end{equation}

If $\frac{\partial\Gamma _{jk}^k}{\partial x^\alpha}=0 $ for
$\delta x^\alpha = \delta x^4, \Gamma _{ik}^k =0$ and $\delta x^
\alpha = 0$ after taking derivatives then

\begin{equation}
 p_i \delta p^i \psi = \frac{1}{2} \hbar^2 g^{i j} R_{ij} \psi.
\label{eq33k}
\end{equation}
and $p_i \delta p^i \psi =0, \delta p_i \delta p^i \psi = 0$.
Using equations (\ref{eq21k}, \ref{eq28k}, \ref{eq33k}), we get
\begin{equation}
g^{ij} \nabla _i \nabla _j \psi =\left\{\frac{1}{4}R-\left(\frac{m
c}{\hbar}\right) ^2 \right\}\psi \label{eq34k}
\end{equation}
Thus, we have obtained Klein-Gordon equation in Schrodinger form.

\section{APPROXIMATE SOLUTIONS.}\noindent

Assuming that the metrics of space-time is almost Galileo's one,
we can rewrite equation (\ref{eq34k}) in single dimensional
approximation for brane as

\begin{equation}
\left\{(g^{11}_0+h^{11})\frac{\partial^2}{\partial
x^2}+2h^{10}\frac{\partial^2}{\partial x
\partial t }+a\right\}\psi=-(g^{00}_0+h^{00})\frac{1}{c^2}\frac{\partial^2\psi}{\partial
t^2}\label{eq39}
\end{equation}
where

\begin{equation}
a=\left(\frac{m c}{\hbar }\right)^2 -\frac14(g^{11}_0+h^{11})
R_{11},
 \label{eq41}
\end{equation}
Taking $ g^{00}_0 = 1, g^{11}_0 = -1, h^{11} = h, h^{10} = h^{00}
= 0,$ we get
\begin{equation}
(1-h)\frac{\partial^2\psi}{\partial x^2}-a
\psi=\frac{1}{c^2}\frac{\partial^2\psi}{\partial t^2}.\label{eq42}
\end{equation}
Looking for the solution in the form of plane wave
\begin{equation}
\psi=A^{i(kx+\omega t)},
 \label{eq46}
\end{equation}
we obtain the following dispersion equation:
\begin{equation}
(1-h)k^2   + a-\frac{\omega^2 }{c^2} =0.
 \label{eq47}
\end{equation}

It has a solution
\begin{equation}
k= \sqrt{\frac{\frac{\omega }{c}-a}{1-h}} .
 \label{eq48}
 \end{equation}
 Assuming that $h$ and $a $ are small, we approximately get
\begin{equation}
k= \frac{\omega }{c }+\frac{1}{2c}\left(\omega h - \frac{a c^2}{2
\omega}\right).
 \label{eq49}
\end{equation}
So, we have for the frequency shift
\begin{equation}
\Delta \omega = \omega -\omega_0 = \frac{1}{2 }\omega h-\frac{1
}{4}\frac{a c^2 }{\omega}, \label{eq50}
\end{equation}
 Using expressions (\ref{eq41}, \ref{eq50}), we get

\begin{equation}
\Delta \omega =  \frac{1}{2 }\omega h+\frac{1 }{4}\frac{c^2
}{\omega}\left(\frac14 R_{11}-\left(\frac{m
c}{\hbar}\right)^2\right). \label{eq51}
\end{equation}

For a photon at $m = R_{11} = 0$ we have
\begin{equation}
\Delta \omega = \frac{1}{2 }\omega h. \label{eq53}
\end{equation}
that is usual gravitational shift.

Also, we get from the formula (\ref{eq49}) the expression for the
group speed of a particle
\begin{equation}
 \frac{\partial\omega}{\partial k} =  \frac{c
}{1-\frac{c^2}{4\omega^2}\left\{\frac14R_{11}-\left(\frac{m
c}{\hbar}\right)^2\right\}}
  \label{eq54}
\end{equation}
For a photon in flat space, we have $
\frac{\partial\omega}{\partial k} =  c$  or ordinary group speed
of light.

\section{SQUARED DIRAC-FOK-IVANENKO EQUATION.}\noindent

We have derived Klein-Gordon equation for a brane. Solutions of
Klein-Gordon equation have global gauge symmetry that is connected
with conservation of chiral current \cite{Mostafazadeh}. Thus,
brane had chiral symmetry after Big Bang. But then, spontaneous
symmetry breaking had occurred and we must use Dirac-Fok-Ivanenko
equation \cite{Ivanenko} after it that do not provide conservation
of chirality for massive particles. We shall compare Klein-Gordon
equation and squared Dirac-Fok-Ivanenko equation. Let's consider
Dirac-Fok-Ivanenko equation
\begin{equation}\label{eqs1}
 i \gamma^i \left(\nabla_i + \Gamma_i \right)\psi = m \psi.
\end{equation}
with Dirac gamma matrixes $\gamma^i$  and spin connection
$\Gamma_i$. Squaring of this equation yields
\begin{equation}\label{eqs2}
 \left(\gamma^i \gamma^j \nabla_i \nabla_j +\left\{ \gamma^i(\nabla_i \gamma^j)+\gamma^i(\Gamma_i \gamma^j)\right\}\nabla_j +  \gamma^i\nabla_i(\gamma^j \Gamma_j ) + (\gamma^i \Gamma_i )(\gamma^j \Gamma_j ) \right)\psi = -m^2 c.
\end{equation}

If

\begin{equation}\label{eqs4}
\frac{1}{2} \left\{ \gamma^i, \gamma^j \right\} = g^{i j},
\end{equation}
\begin{equation}\label{eqs3}
 \gamma^i (\partial_i\gamma^j)+ \{(\gamma^i \Gamma_i ),\gamma^j\}= 0,
\end{equation}
and
\begin{equation}\label{eqs5}
\gamma^i \nabla_i(\gamma^j \Gamma_j)+(\gamma^i \Gamma_i)(\gamma^j
\Gamma_j)= -\frac{1}{4}R,
\end{equation}
we come Klein-Gordon equation (\ref{eq39}).

    Relations (\ref{eqs4},\ref{eqs3} ) coincide with that of \cite{Jackiw}. Taking into account expression from \cite{Weldon}
\begin{equation}\label{eqs6}
\partial_i \gamma^j+\left[ \Gamma_i ,\gamma^j\right] +\Gamma ^i_{jk}\gamma^k= 0,
\end{equation}

    we come as in \cite{Jackiw} from (\ref{eqs4}) and (\ref{eqs3}) to the following relation between spin connection and Cristoffel symbol
\begin{equation}\label{eqs7}
\Gamma^j =\frac{1}{2}g^{ik}\Gamma ^j_{ik},
\end{equation}

    Let's consider relation (\ref{eqs5}). Using (\ref{eqs6}) again we get
\begin{equation}\label{eqs8}
\gamma^i \gamma^j \nabla_i \Gamma_j +\gamma^i \gamma^j
\Gamma_i\Gamma_j = -\frac{1}{4}R,
\end{equation}
    where $\nabla_i \Gamma_j= \partial_i\Gamma_j - \Gamma ^k_{ij} \Gamma_k$  is covariant derivative. Therefore we obtain that
\begin{equation}\label{eqs9}
\gamma^i \gamma^j D_i \Gamma_j = -\frac{1}{4}R,
\end{equation}
where $D_i = \nabla_i+ \Gamma_i $  is generalized covariant
derivative.

\section{Chiral current.}\noindent

Let's consider the first term in (\ref{eqs9}). It can be
decomposed into commutator and anticomutator parts
\begin{equation}\label{eq14}
  \gamma^i \gamma^j D_i \Gamma_j =\frac{1}{2}\{\gamma^i ,\gamma^j\}D_i \Gamma_j +
  \frac{1}{2}[\gamma^i ,\gamma^j]\left(\frac{1}{2}(D_i \Gamma_j + D_j \Gamma_i)+\frac{1}{2}(D_i \Gamma_j - D_j \Gamma_i) \right).
\end{equation}

It can be rewritten as
\begin{equation}\label{eq15}
  \gamma^i \gamma^j D_i \Gamma_j =g^{i\nu} D_i \Gamma_j +
  \frac{1}{4}[\gamma^i,\gamma^j]\left(\nabla_i \Gamma_j - \nabla_j \Gamma_i+\Gamma_i \Gamma_j - \Gamma_j \Gamma_i \right).
\end{equation}

Introducing spin curvature
\begin{equation}\label{eq16}
\Phi_{ij} =\nabla_i \Gamma_j - \nabla_j \Gamma_i+\Gamma_i \Gamma_j
- \Gamma_j \Gamma_i = D_i\Gamma_j -D_j\Gamma_i
\end{equation}
we get from (\ref{eqs9}) the expression
\begin{eqnarray}\label{eq17}
[\gamma^i ,\gamma^j]\Phi_{ij} &=&  -4 g^{ij} D_i \Gamma_j - R.
\end{eqnarray}
that can be used for determination of $[\gamma^i ,\gamma^j]$.

Another variant of this formula is derived as following. Spin
curvature can be expressed as \cite{Weldon}

\begin{eqnarray}
\Phi_{ij} &=&-\frac{1}{8}  [\gamma^l ,\gamma^k] R_{ijlk}.
\label{eq18}
\end{eqnarray}

Substitution of (\ref{eq18}) into (\ref{eq17}) gives
\begin{equation}
\left(\frac{1}{8}[\gamma^i ,\gamma^j][\gamma^l ,\gamma^k]-
\frac{1}{4}\{\gamma^j ,\gamma^k\}\{\gamma^i
,\gamma^l\}\right)R_{ijlk}=4 g^{ij} D_i \Gamma_j, \label{eq19}
\end{equation}
Using symmetry properties of metric tensor and Bianci identity we
get
\begin{equation}
\gamma^i \gamma^j \gamma^k \gamma^l R_{iljk} = 4 g^{ij} D_i
\Gamma_j, \label{eq20}
\end{equation}
The left hand side of (\ref{eq20}) can be rewritten as
\begin{equation}
\gamma^i \gamma^j \gamma^k \gamma^l R_{iljk} =
\delta^{ijps}_{klmn} \gamma^k \gamma^l \gamma^m \gamma^n R_{isjp}.
\label{eq21}
\end{equation}

where tensor $\delta^{ijps}_{klmn}$ is the generalized Kronecker
symbol. Further, we use the identity
\begin{equation}
\delta^{ijps}_{klmn}
=\frac{1}{4!}\varepsilon^{ijps}\varepsilon_{klmn}. \label{eq22}
\end{equation}
substituting it in (\ref{eq21}) to get
\begin{equation}
\gamma^i \gamma^j \gamma^k \gamma^l R_{iljk}  = -i \gamma^5
\widetilde{R}, \label{eq23}
\end{equation}
where
\begin{equation}
\gamma^5 =\frac{i}{4!}\varepsilon_{klmn}\gamma^k\gamma^l \gamma^m
\gamma^n. \label{eq24}
\end{equation}
and
\begin{equation}
\widetilde{R} = \varepsilon^{ijps} R_{isjp}, \label{eq25}
\end{equation}
Eventually, we get from (\ref{eq20}) and (\ref{eq23}) the
following expression for :
\begin{equation}
\gamma^5  = 4i\widetilde{R}^{-1} g^{ij} D_i \Gamma_j.\label{eq26}
\end{equation}
or
\begin{equation}
\gamma^5  = 4i\widetilde{R}^{-1}  D_i \Gamma^j.\label{eq26o}
\end{equation}
 $\gamma^5$ can be used for the determination of chiral spin current according to the formula
\begin{equation}
\gamma^{k 5}  = \overline{\psi}\gamma^{k}\gamma^{ 5}
\psi.\label{eq27}
\end{equation}
 where $\psi$  is spinor wave function. Substitution of (\ref{eq26}) into (\ref{eq27}) yields
\begin{equation}
j^{k 5}  = 4i\overline{\psi}\gamma^{k}\widetilde{R}^{-1} D_i
\Gamma^i \psi.\label{eq28}
\end{equation}

Compare it with the expression for topologically nontrivial vacuum
current of abnormal parity
\begin{equation}
j^{k}  = \frac{1}{8\pi}e c \varepsilon^{kij}F_{ij} .\label{eq29}
\end{equation}
where $F_{ij}$  is electromagnetic field strength, in the theory
of fractional Hall effect \cite{Jackiw}. Both formulas are
relativistic, and in both cases current emerge due spin-orbit
interaction in nontrivial field topology. The difference is that
whereas, in the second case, we deal with a magnetic field, in the
first case, we have gravitation affine gauge field
\begin{equation}
G_{ij}  = \nabla_s \Gamma^{s}_{ij}
+\frac{1}{2}g_{sp}\Gamma^{s}_{ij}
g^{i'j'}\Gamma^{p}_{i'j'}.\label{eq30}
\end{equation}
corresponding by its structure to that of works \cite{Hecht, Hehl}
 and yielding the chiral current
\begin{equation}
j^{k 5}  =
i\overline{\psi}\gamma^{k}\widetilde{R}^{-1}g^{ij}(G_{ij}-G_{ji})
 \psi.\label{eq31}
\end{equation}
in its symmetric form.


\section{Conclusion}\noindent

Thus, we have derived Klein-Gordon equation for a particle on
brane using variation principle. Solutions of Klein-Gordon
equation have chiral symmetry and thus brane after its emergence
due Big Bang supported chiral symmetry. But then this symmetry was
spontaneously broken and we must deal with Dirac equation. It can
be verified that the Dirac decomposition of obtained Klein-Gordon
equation yields Dirac-Fock-Ivanenko equation that can be solved
with Hilbert-Einstein equation. Squaring Dirac-Fock-Ivanenko
equation gives wave equation \cite{Mostafazadeh},
\cite{Shavokhina}, \cite{Metlitski} coinciding with that obtained
in the present paper.

Solution of modified Dirac-Fock-Ivanenko equation for a photon on
the background of almost Galileo's metrics yields the changes in
frequency and speed of particle's wave packet that can be verified
experimentally.

We have derived the expression for gamma five matrix through the
derivative of spin connection by comparison of squared DFI
equation with KG equation and used it for determination of chiral
spin current. This chiral spin current is anomalous spin current
corresponding to spontaneous chiral symmetry breaking of mass
particle in the space of KG equation solutions. Derived formula
for chiral spin current in affine gravitational  gauge field has
the structure analogous to that of anomalous Hall current
\cite{Jackiw} and topological axial current in dense matter
induced by magnetic flux \cite{Metlitski}.

\end{document}